\documentclass{article}

\usepackage{graphicx}

\begin{document}

\title{Development of a Novel Quantum Pre-processing Filter to Improve Image Classification Accuracy of Neural Network Models}

\maketitle

\author{Farina Riaz, Shahab Abdulla, Hajime Suzuki, Srinjoy Ganguly, Ravinesh C.~Deo, and Susan Hopkins}

%
%


\begin{abstract}

This paper proposes a novel quantum pre-processing filter (QPF) to improve the image classification accuracy of neural network (NN) models. A simple four qubit quantum circuit that uses Y rotation gates for encoding and two controlled NOT gates for creating correlation among the qubits is applied as a feature extraction filter prior to passing data into the fully connected NN architecture. By applying the QPF approach, the results show that the image classification accuracy based on the MNIST (handwritten 10 digits) and the EMNIST (handwritten 47 class digits and letters) datasets can be improved, from 92.5\% to 95.4\% and from 68.9\% to 75.9\%, respectively. These improvements were obtained without introducing extra model parameters or optimizations in the machine learning process. However, tests performed on the developed QPF approach against a relatively complex GTSRB dataset with 43 distinct class real-life traffic sign images showed a degradation in the classification accuracy. Considering this result, further research into the understanding and the design of a more suitable quantum circuit approach for image classification neural networks could be explored utilizing the baseline method proposed in this paper.

\end{abstract}

\section{Introduction}

The application of quantum computing to the tasks of machine learning, herein referred as quantum machine learning, has attracted much research attention in recent years. Literature surveys on quantum machine learning can be found in \cite{3844,3824,3935}. Among many proposals to combine classical machine learning methods with quantum computing, quanvolutional neural network (QNN) proposed by Henderson et.\ al.\ \cite{3882} has an advantage of being implementable on quantum circuits with a smaller number of qubits with shallow gate depths and yet being applicable to practical applications. QNN utilizes quantum circuits as transformation layers, called quanvolutional layer, to extract features for the purpose of image classification using convolutional neural networks (CNNs). In \cite{3882}, MNIST handwritten 10 digit dataset \cite{3937} was applied to QNN using 9 qubits. The results showed classification accuracy improvement using QNN over CNN. However, when the quanvolutional layer of QNN was replaced by a conventional convolutional layer, no improvement in classification accuracy was observed. Henderson et.\ al.\ later updated QNN and implemented on Rigetti Computing’s Aspen-7-25Q-B quantum processing unit which has 25 qubits with 24 programmable two-qubit gate interactions \cite{3960}. The proposed method was applied to 4 class low resolution satellite image dataset. However, no improvement in classification accuracy by QNN over CNN was observed in \cite{3960}.

An implementation of QNN on a software quantum computing simulator, PennyLane \cite{3664}, was provided by Mari \cite{4013}. Mari's implementation of QNN differs from that of Henderson in that the output of the quantum circuit, which is a set of expectation values, is directly fed into the following neural network layer, while that of Henderson was made into a single scalar value by a classical method. The proposed method was applied to MNIST dataset using 50 training and 30 test image set. No clear improvement in classification accuracy by QNN over NN was observed in \cite{4013}.

Inspired by Henderson's and Mari's QNNs, we propose a new method, herein called quantum pre-processing filter (QPF), that shows clear improvements in image classification accuracy over conventional neural networks. QPF uses a quantum circuit with four qubits, four Y rotations, two controlled NOTs (CNOTs), and four measurements. When QPF is applied as a pre-processing unit of an image classification neural network, i.e.\ as a feature extraction filter, the image classification accuracy of fully connected neural network against MNIST and EMNIST (handwritten 47 class digits and letters, \cite{3968}) improves from 92.5\% to 95.4\% and from 68.9\% to 75.9\%, respectively. These improvements were obtained without introducing any extra parameters to optimize in machine learning process. Unlike other quantum machine learning methods, the use of QPF does not require optimization of the parameters inside the quantum circuits and hence requires only a limited use of the quantum circuit. Given the small number of qubits and relatively shallow depth of the quantum circuit, QPF is well suited to be implemented on noisy intermediate-scale quantum computers. While the proposed method is promising, a test against a more complicated dataset, GTSRB (43 class real-life traffic sign images, \cite{3847}), showed degradation in classification accuracy by the application of QPF. This prompts further research into the understanding and design of suitable quantum circuits for image classification neural networks.  To support the validation of our claims and further research, we have made our source code available at \verb+https://github.com/hajimesuzuki999/qpf+.

This paper is organized as follows: The new QPF unit combined with the classical image classification neural network is proposed in Section 2. Section 3 describes the experiment conducted using software simulation. The results and discussions are presented in Section 4, followed by conclusions in Section 5.

\section{Methodology}

\begin{figure}[h]
  \includegraphics[width=\linewidth]{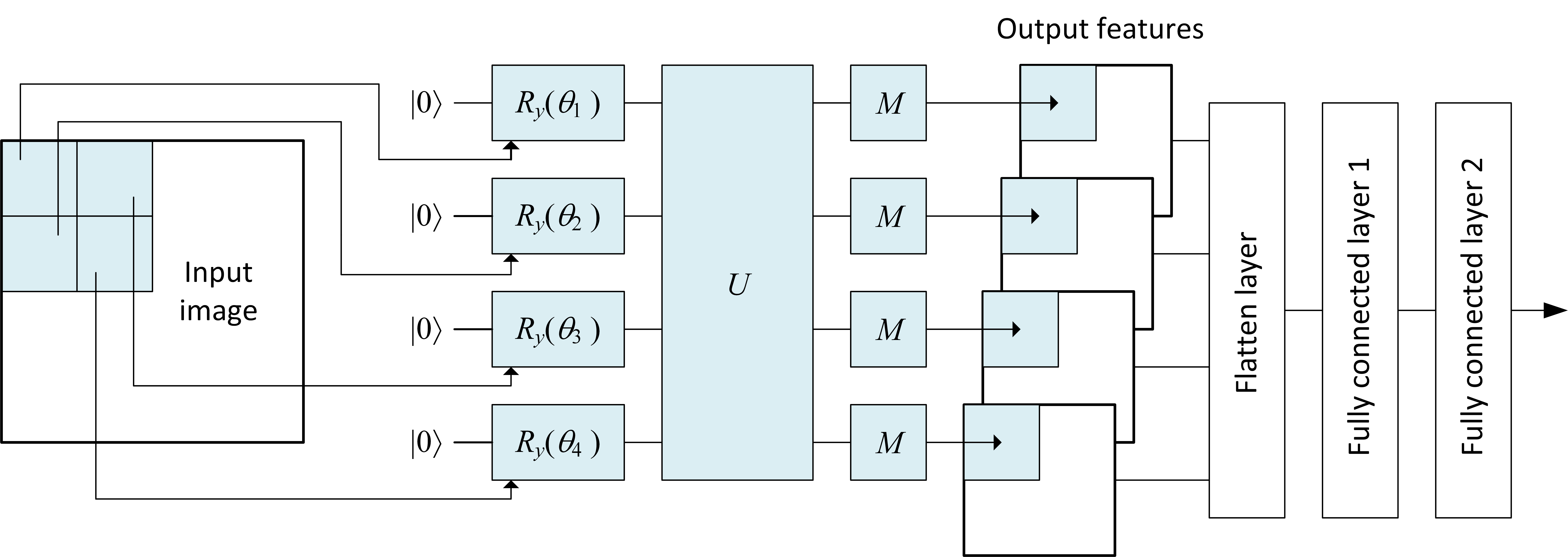}
  \caption{The architecture of the proposed quantum pre-processing filter (QPF) model.}
  \label{fig:architecture}
\end{figure}

Figure~\ref{fig:architecture} shows the architecture of the proposed model. The method assumes that the input image is a two-dimensional matrix with size $m$-by-$m$ and the pixel value, $x$, follows $0 \leq x \leq 1$. An extension to multi-channel pixel image is considered as straightforward. Similar to QNN models, a section of size $n$-by-$n$ is extracted from the input image. While $1 < n \leq m$ in the case of QNN, the proposed QPF uses $n = 2$. This $2 \times 2$ section of the input image is referred as QPF window. An extension of QPF using $n > 2$ is left for further studies.

Given $n = 2$, we use 4 qubit quantum circuit as shown in Figure~\ref{fig:architecture}. The four qubits are initialized in the ground state. The four pixel values are then encoded using Y rotation with $\theta = \pi x$ according to (1).

\begin{equation}
R_y(\theta) = \left[ \begin{array}{cc}

\cos \frac{ \theta }{ 2 } & - i \sin \frac{ \theta }{ 2 } \\
\sin \frac{ \theta }{ 2 } & \cos \frac{ \theta }{ 2 }

\end{array} \right]
\label{eqn:y_rotation}
\end{equation}

\begin{figure}
\centering
  \includegraphics[scale=0.8]{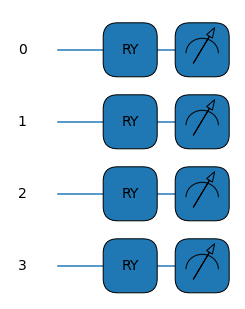}
  \caption{Quantum pre-processing filter, with encoding only.}
  \label{fig:encoding_only}
\end{figure}

\begin{figure}
\centering
  \includegraphics[scale=0.8]{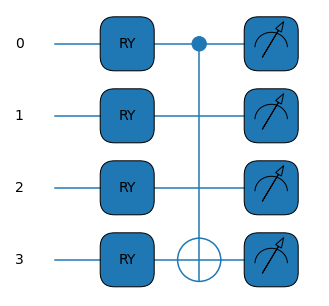}
  \caption{Quantum pre-processing filter, with one CNOT.}
  \label{fig:one_cnot}
\end{figure}

\begin{figure}
\centering
  \includegraphics[scale=0.8]{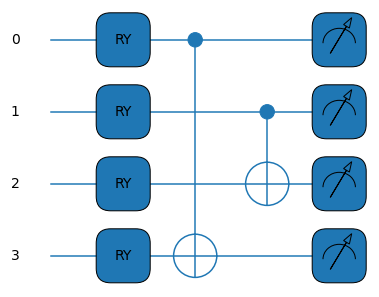}
  \caption{Quantum pre-processing filter, with two CNOTs.}
  \label{fig:two_cnots}
\end{figure}

The outputs from the Y rotation gates are fed to the quantum circuit referred as $U$ in Figure~\ref{fig:architecture}. Measurements, referred as $M$ in Figure~\ref{fig:architecture}, are performed on the output of the quantum circuit $U$. Three different quantum circuits are examined in this paper. The first circuit, referred as Encoding only and shown in Figure~\ref{fig:encoding_only}, performs measurement straight after the encoding.  The second circuit, referred as One CNOT and shown in Figure~\ref{fig:one_cnot}, performs controlled NOT operation with the first qubit as the control and the 4th qubit as the target, as shown in Figure~\ref{fig:one_cnot}.  The third circuit, referred as Two CNOTs, performs two controlled NOT operations as shown in Figure~\ref{fig:two_cnots}.

The outputs from the measurement operations are given as expectation values between $-1$ and $1$, and form output features. We note that the total number of parameters in the input image ( $m \times m$ ) is the same as the total number of parameters in the output features ( $4 \times ( m / 2 ) \times ( m / 2 )$ ). The output features are made into a one dimensional vector by the flatten layer.  The number of the nodes of the output of the flatten layer is $m \times m$.  The nodes are fully connected by the first fully connected layer 1.  The output of the fully connected layer 2 has the number of nodes equal to the number of classes.


\section{Experiment}

As was performed by many, we first apply the proposed method to the MNIST dataset \cite{3937} to obtain benchmark results. The MNIST dataset consists of 60,000 training and 10,000 test images of handwritten digits of 0 to 9. The size of each image is 28 by 28 pixels. The original images are in greyscale within the values between 0 and 255, which are scaled by dividing them by 255. We then chose the EMNIST dataset \cite{3968} to extend the number of image class. The EMNIST dataset (Balanced and By\_Merge \cite{3968}) contains 112,800 training and 18,800 test images of handwritten digits and letters making up 47 classes. Note that some of upper- and lower-case letters are merged due to their similarity (e.g. C is similar to c) in this dataset. Original EMNIST dataset is divided by 255 to create a dataset with pixel values between 0 and 1.

The GTSRB dataset \cite{3847} consists of 34,799 training and 12,630 test images of 43 class traffic signs captured from actual traffic signs in-use in various conditions. The original dataset has various image sizes between $15 \times 15$ and $222 \times 193$ pixels. Those images were scaled to a size of $32 \times 32$ pixels. The original images were in RGB color, which were converted into grayscale between 0 and 255. Unlike MNIST and EMNIST dataset, in order to normalize the dynamic range of each image, the normalization is applied to each image according to the following formula:

\begin{equation}
\tilde{c}_{x,y} = \frac{ c_{x,y} - \min( I ) }{ \max( I ) - \min( I ) }
\label{eqn:scaling}
\end{equation}

where $c_{x,y}$ and $\tilde{c}_{x,y}$ represent the original and normalized pixel values in the position ($x$,$y$), and $\max(I)$ and $\min(I)$ denote the maximum and the minimum element in the two dimensional image matrix $I$, respectively.

The experiment was performed mainly in MATLAB.  PennyLane was used to build quantum circuits which were exported to MATLAB as unitary matrices.  The Adam optimizer \cite{4014} was used as the solver for the training network and default mini-batch size of 128 was used for all of the three datasets to fully test the proposed QPF model.

\begin{figure}
  \includegraphics[width=\linewidth]{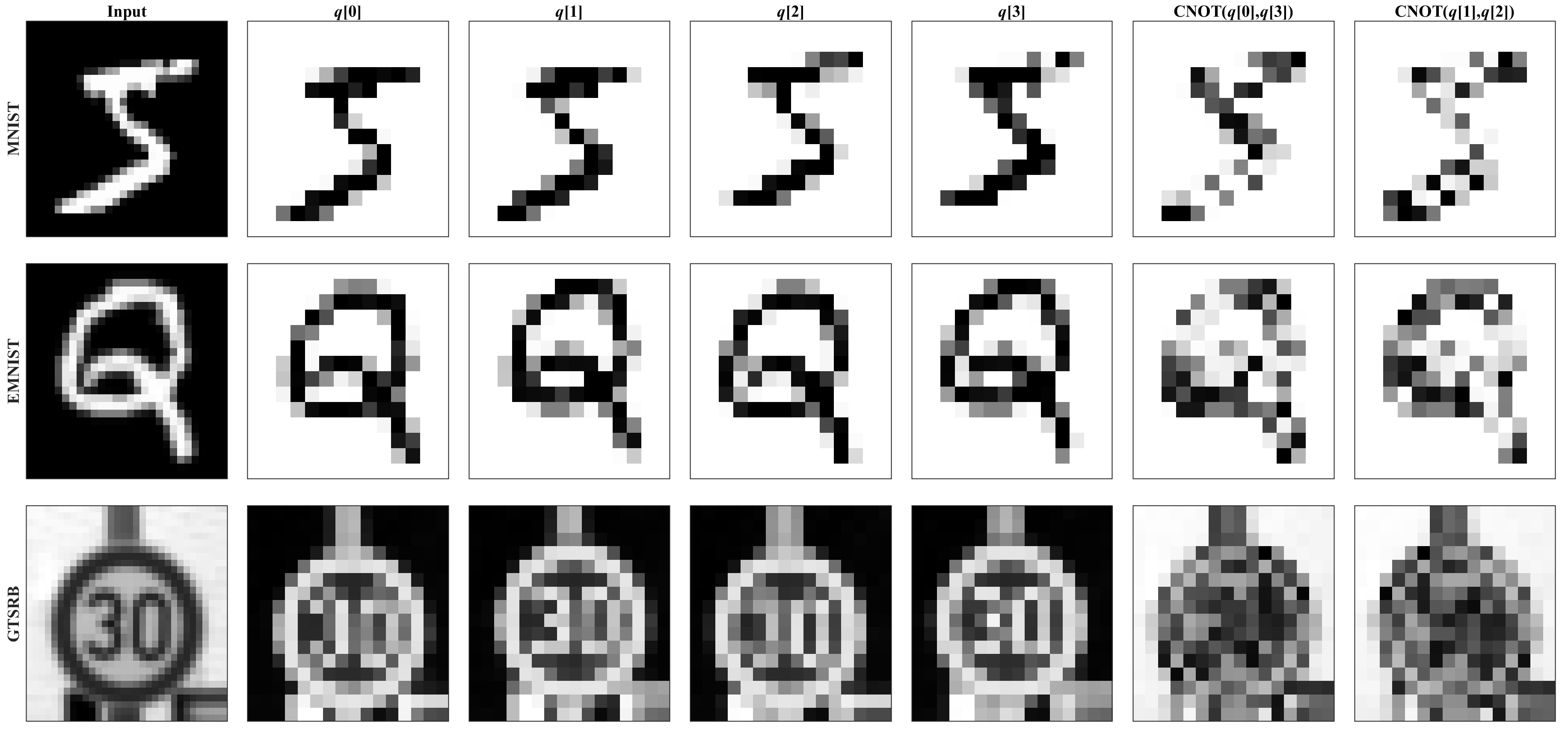}
  \caption{An example of the input image and the relevant output features.}
  \label{fig:feature}
\end{figure}

Figure~\ref{fig:feature} shows example input images from MNIST, EMNIST, and GTSRB, and corresponding output features using encoding only (labelled as $q[0]$ to $q[3]$), and two CNOTs (CNOT( $q[0],q[3]$ ) and CNOT( $q[1],q[2]$ )).

\section{Results and Discussion}

\begin{figure}
\centering
  \includegraphics[scale=0.8]{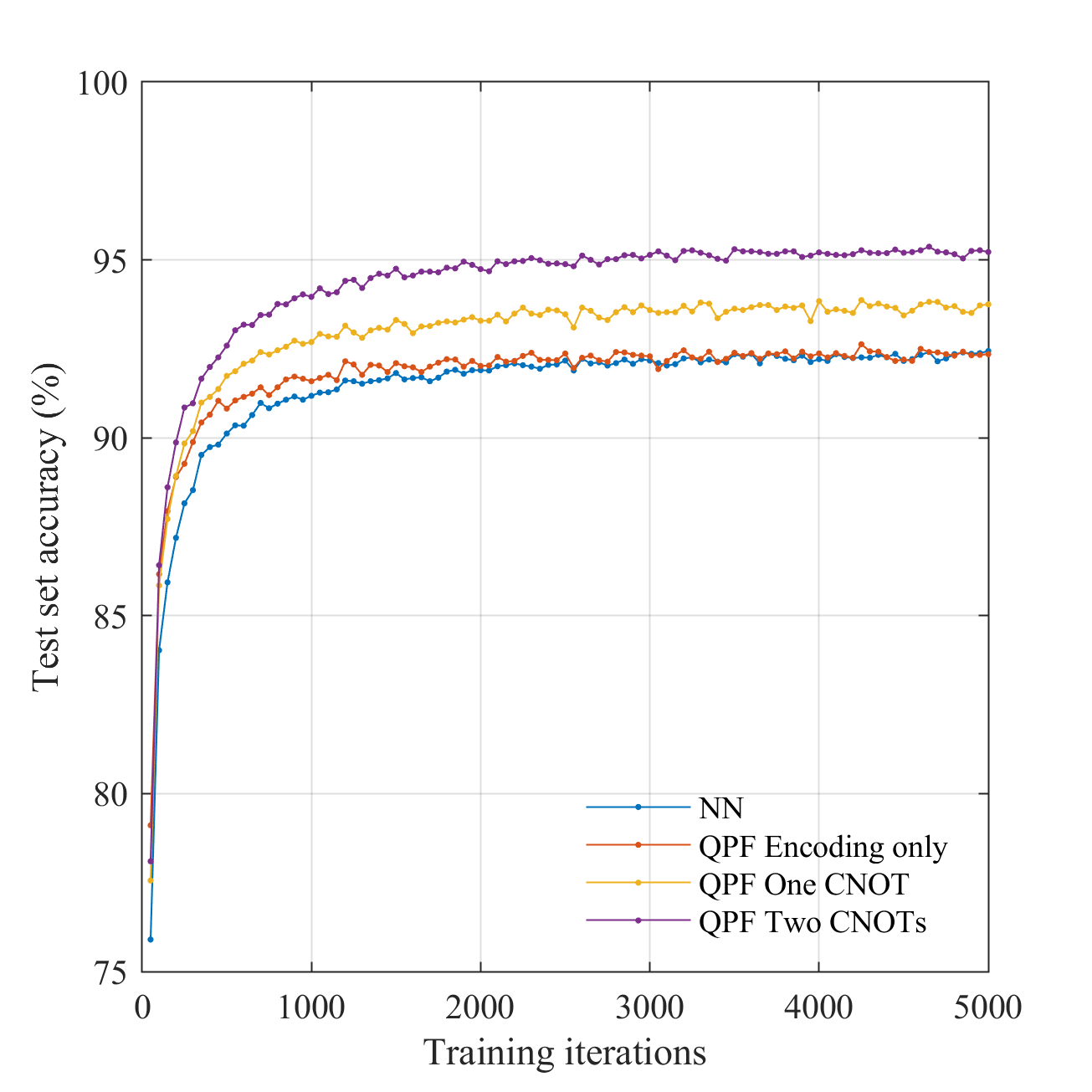}
  \caption{Test set accuracy using MNIST dataset.}
  \label{fig:mnist}
\end{figure}

Figure~\ref{fig:mnist} shows the variation of the testing dataset accuracy as a function of training iterations using MNIST dataset. As can be seen in Figure~\ref{fig:mnist}, the test set accuracy using QPF Encoding only converges faster than that of NN. The phenomenon of faster convergence was also observed in Henderson’s QNN \cite{3882}. However, the converged test set accuracy of the QPF Encoding only model does not improve that of NN.  The application of QPF One CNOT improves the test set accuracy from 92.5\% to 93.7\%.  The test set accuracy is further improved to 95.4\% by the application of QPF Two CNOTs.

\begin{figure}
\centering
  \includegraphics[scale=0.8]{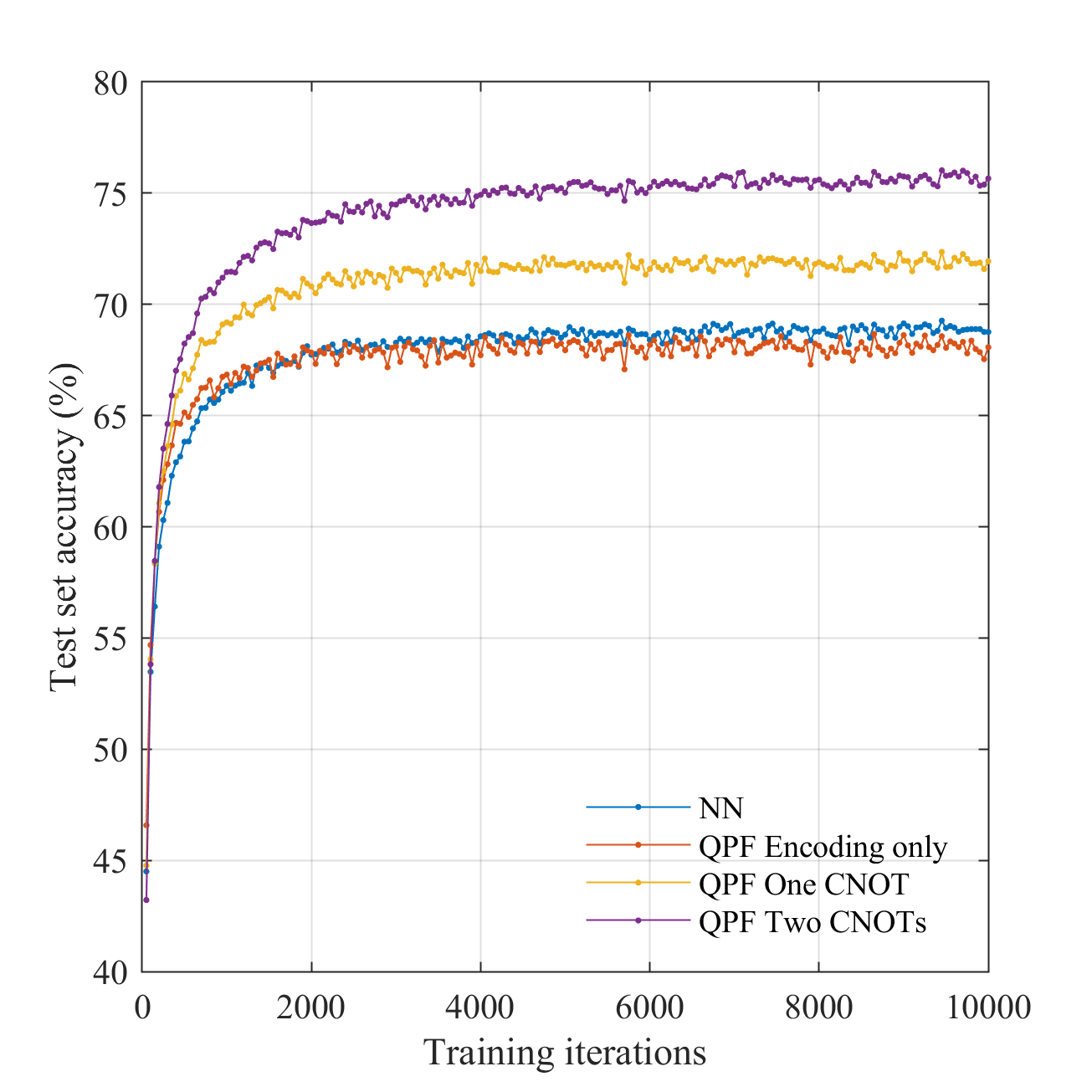}
  \caption{Test set accuracy using EMNIST dataset.}
  \label{fig:emnist}
\end{figure}

Similarly, small but clearly faster convergence is observed by the application of QPF Encoding only in the case of EMNIST dataset as shown in Figure~\ref{fig:emnist}. However, in the case of EMNIST, the converged test set accuracy of the QPF Encoding only is reduced from that of NN. Nonetheless, the test set accuracy is improved from 68.9\% to 72.0\% by the application of QPF One CNOT, and to 75.9\% by the application of QPF Two CNOTs.

\begin{figure}
\centering
  \includegraphics[scale=0.8]{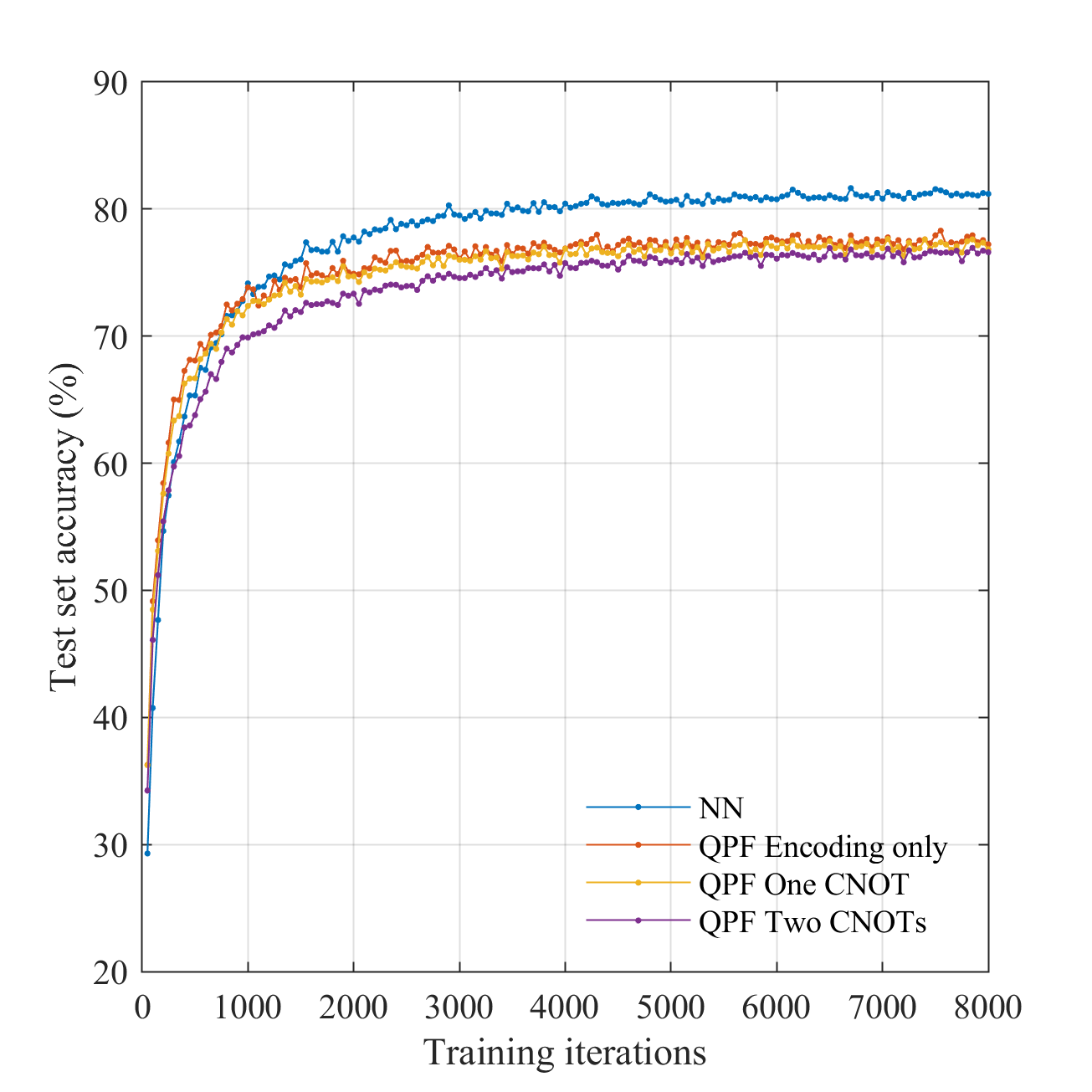}
  \caption{Test set accuracy using GTSRB dataset.}
  \label{fig:gtsrb}
\end{figure}

Figure~\ref{fig:gtsrb} shows the variation of test set accuracy using the GTSRB dataset. Unlike the cases using MNIST and EMNIST datasets, the converged test set accuracy by the application of QPF is reduced from that of NN in the case of GTSRB dataset. The exact causes of this phenomenon are currently unknown to the authors and remain for further research.  The summary of the testing accuracy results is shown in Table~\ref{tbl:summary}.

\begin{table}
\centering
\begin{tabular}{c|ccc}
& MNIST & EMNIST & GTSRB \\ \hline
NN & 92.5\% & 68.9\% & 81.4\% \\
Encoding only & 92.4\% & 68.1\% & 77.6\% \\
One CNOT & 93.7\% & 72.0\% & 77.6\% \\
Two CNOT & 95.4\% & 75.9\% & 77.0\% \\ \hline
\end{tabular}
\caption{Summary of testing accuracy results.}
\label{tbl:summary}
\end{table}

\begin{figure}
\centering
  \includegraphics[scale=0.8]{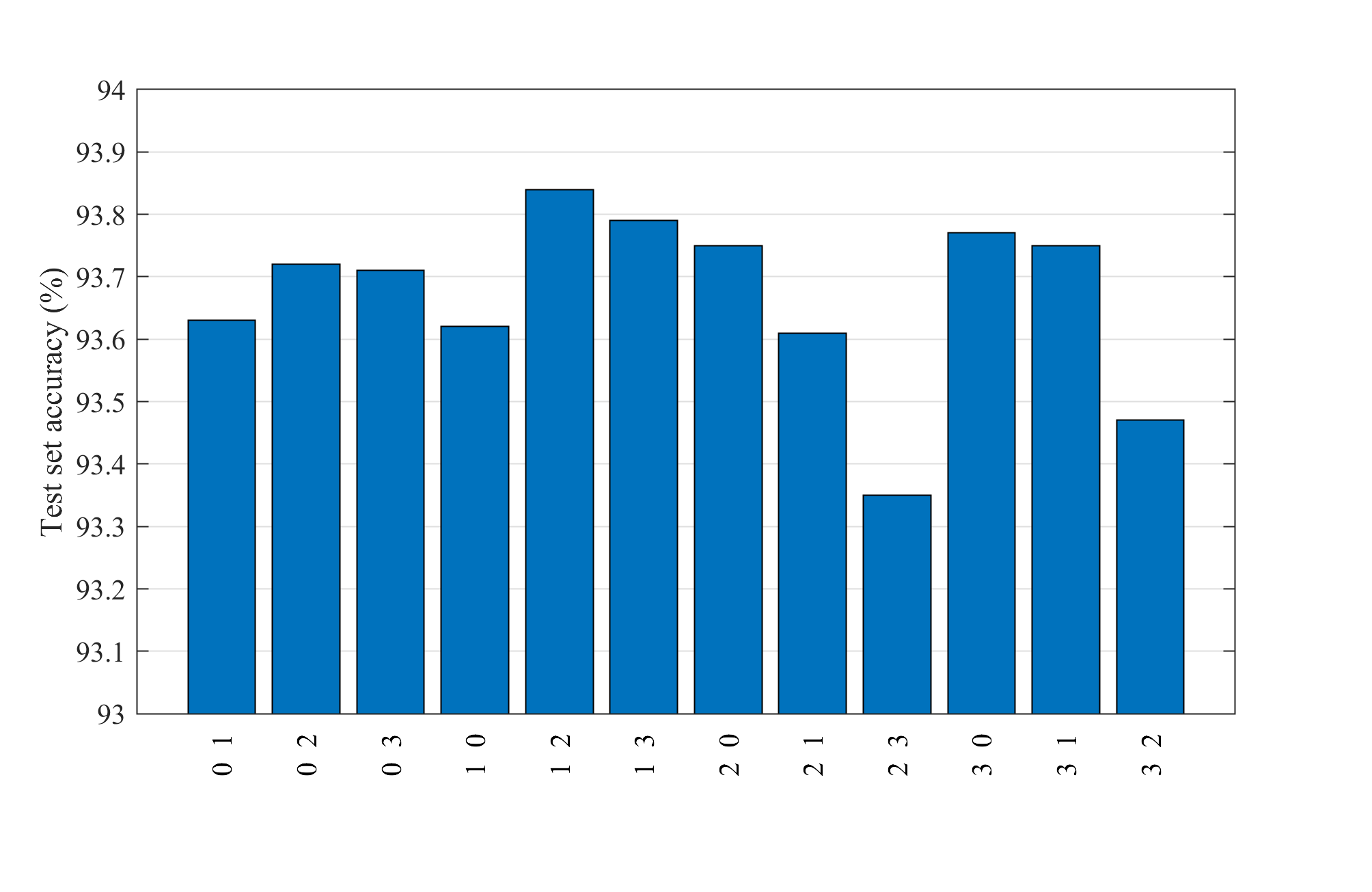}
  \caption{Test set accuracy on MNIST dataset using different CNOT arrangements.}
  \label{fig:one_cnot_all}
\end{figure}

Referring back to Figure~\ref{fig:one_cnot}, there are 12 different ways to create a CNOT circuit from four qubits. In order to investigate if a different CNOT arrangement would make the difference in classification accuracy, the training of the network and the classification of the images were performed for MNIST dataset using the 12 different CNOT arrangements. Figure~\ref{fig:one_cnot_all} shows the results. The $x$ axis of Figure~\ref{fig:one_cnot_all} denotes the arrangement of the CNOT where the lower number refers to the control qubit and the upper number refers to the target qubit. For example, “0 3” refers to the arrangement as shown in Figure~\ref{fig:one_cnot}. As can be seen from Figure~\ref{fig:one_cnot_all}, the variation of test set accuracy for different CNOT arrangement is considered to be small, within less than 0.6\%.

\begin{figure}
\centering
  \includegraphics[scale=0.8]{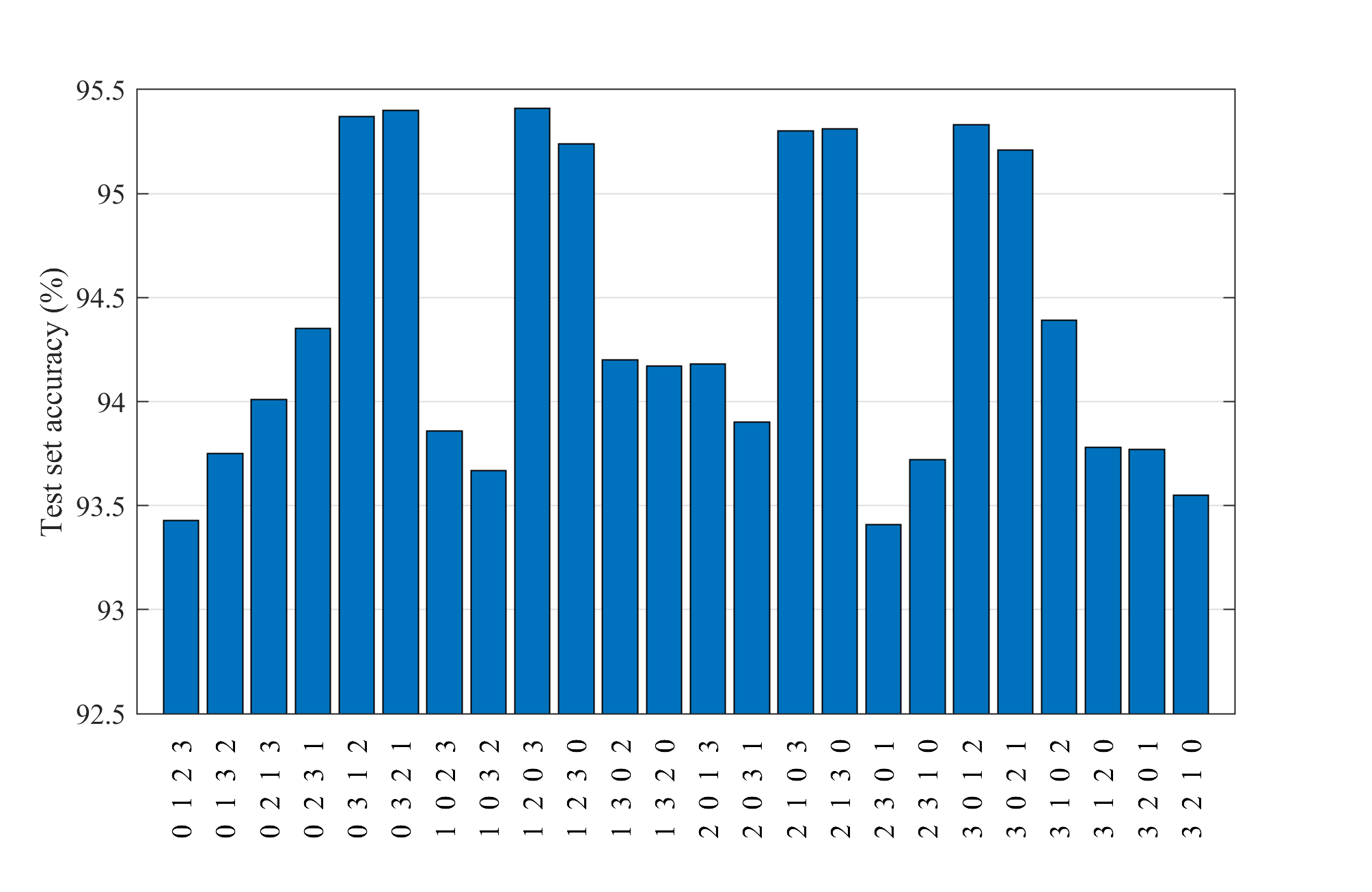}
  \caption{Test set accuracy on MNIST dataset using different two CNOTs arrangements.}
  \label{fig:two_cnot_all}
\end{figure}

Similarly, there are 24 different ways to arrange the two CNOTs using four qubits. The test set accuracy was derived for 24 different two CNOTs arrangement using MNIST dataset, and the results are shown in Figure~\ref{fig:two_cnot_all}. In Figure~\ref{fig:two_cnot_all}, the $x$ axis label refers to the arrangement of the two CNOTs in the order of the control and target qubits of the first CNOT, then the control and target qubits of the second CNOT. For example, “0 3 1 2” refers to the arrangement as shown in Figure~\ref{fig:two_cnots}. As evident in Figure~\ref{fig:two_cnot_all}, the CNOT arrangements pairing the 1st and the 4th qubits and pairing the 2nd and the 3rd qubits seems to achieve a higher testing accuracy irrespective to which qubit is assigned as the target or the control. Referring back to Figure~\ref{fig:architecture}, the pairing of the 1st and 4th qubits and the pairing of the 2nd and 3rd qubits correspond to the pairing the diagonal elements of the 2 x 2 QPF window. The exact reason for the improved classification accuracy in the case of MNIST and EMNIST dataset when correlating the diagonal elements of the QPF window is currently not known to the authors and is left for further research.

\section{Conclusion}

A novel QPF that can improve the image classification accuracy of NN for MNIST and EMNIST datasets is proposed. While clear improvements are observed, the exact causes of the improvements are currently unknown and remain for future research and investigation.  Also, further investigation is needed in order to improve the classification accuracy of NNs against complex images such as those from the GTSRB dataset.  Proposed future research directions include increasing the number of qubit sizes from four as well as applying the QPFs to other image classification methods such as convolutional NNs.

\section{Acknowledgment}

This research has been supported by Australian government Research Training Program and Commonwealth Scientific Industrial and Research Organization.

\bibliographystyle{plain}
\bibliography{IEEEabrv,hajimesbib}

\end{document}